# Determinants of Lending to Small and Medium Enterprises by Commercial Banks in Kenya

Shikumo, David Haritone[1] &Mwangi, Mirie[2]

[1-]Master of Science in Finance Graduate- University of Nairobi, Kenya
[2-]Senior Lecturer-Department of Finance and Accounting, University of Nairobi, Kenya

***Abstract:*** *Small and Medium Enterprises (SMEs) access to external finance is an issue of significant research interestto academicians. Commercial banks consider many SMEs not to be credit worthy because of their inability to meet some banking requirements. Hence, the objective of this study was to investigate what determines lending to SMEs by commercial banks in Kenya. To achieve the study objectives, a descriptive research design was employed. The study undertook a census of the 43 commercial banks in Kenya, with full data being obtained for 36 institutions. The study used secondary data from the annual published reports of commercial banks in Kenya for a period of 5 years from 2010-2014. The data collected was analyzed through the multiple linear regression using the Statistical Package for Social Studies version 20.The study established that bank size and liquidity significantly influences (positively and negatively, respectively) lending to SMEs by commercial banks in Kenya while credit risk and interest rates have no significant influence on lending to SMEs by commercial banks in Kenya. The study recommends that lending to SMEs by commercial banks in Kenya be enhanced by adopting policies that grow the commercial banks.*

## I. Introduction

Commercial banks play a significant role in economic resource allocation of many countries around the globe. They channel funds from depositors to investorsas well as generating the necessary income to cover their operational costs (Ongore & Kusa, 2013).Commercial banks' ending significantly plays crucial roles in catalyzing industrialization in the economy, by facilitating the mobilization and transfer of funds for economic production. Therefore, well-functioningcommercial banks spur technological innovations by identifying and funding entrepreneurs, thus creating chances of successfully implementing innovative products and production process (Olusanya, Oyebo & Ohadebere, 2012). One of the objectives of commercial bank lending is to improve private sector business activity, to enhance their contribution to economic growth. However, commercial banks have basic lending principles that act as a check on their lending activities.

In most of the developing economies, commercial banks are often unable or reluctant to grantloans to small and medium enterprises. Instead, they prefer lending to wellestablishedbusinesses that have well maintainedfinancial statements and credit histories. The formal financing to SMEs is mostly obstructed by the collateral requirement in conventional banking. This can be attributed to the SMEs size and age, lack of business strategy, collateral, financial information, bank requirements as well as the owners or manager's educational background and business experience (Abdesamed&Wahab, 2014). Commercial banks prefer to lend to businesses with proper financial statements or recordsas well as sufficient collateral in form of tangible assets, which are difficult for SMEs to obtain. SMEs also suffer financing shortage due to information asymmetry and their inaccessibility to debt finance, forces SMEs to use internal capital, whichmay beinsufficient for expansion.

**1.1.1 Lending to SMEs**

Lending is one of the major activities from which commercial banks earn their profit (Asantey&Tengey, 2014). Commercial bank lending is an important source of finance to many businesses especially the SMEs, which are more reliant on traditional debt to fulfill their business financial needs (OECD, 2015).However, lending to SMEs by commercial banks poses the most serious credit risks. Credit risk constitutes the likelihood that the SME would default on interest and/or principal (Obamuyi, 2007). Credit risk is a major concern to all financial institutions that are involved in lending to SMEs because the risk of default by SME clients can jeopardize the performance and survival of the lending institution (Ojiambo, 2012).

Commercial banks that are supposed to finance SMEs usually face lack of accurate and reliable information on the financial condition and performance of SMEs. Thus, in order to extend credit to SMEs, the commercial banks' lending decision is based on the strength of audited financial statements, long credit histories andgood history of the principal owner (Chepkorir, 2013), most of which SMEs do not have.Despite the fact that financial institutions have identified SMEs as fast growing entities, there are several constraints serving as bottlenecks to SMEs in accessing finance from financial institutions (Kwaning et al., 2015). As such, access to traditional bank finance by SMEs is a great challenge especially to newer, innovative and fast growing businesses, with a higher risk-return profile (OECD, 2015).

Lending to SMEs is perceived to be more risky than to large established firms (Hall and Fang,2004), thus, the requirement of collateral in form of tangible assets by most of financial institutions and non-financial institutions in addition to other stringent requirements to reducerepayment default rate. Commercial banks financing SMEs also face some difficulties in obtaining accurate and reliable information on the performance as well as the financial condition of the SMEs. As such, Commercial banks tend to hesitate in financing startups especially those with insufficient collateral (Kravchenko, 2011). In addition, most commercial banks tend to monitor the firmsthey lend funds so as to ensure that such funds are well utilized as per the initial contract and purpose. It becomes difficult to monitor SMEs than large established firms hence commercial banks are more likely to engage in credit rationing to SMEs than to large establishedfirms (Alhassan and Sakara, 2014).

### 1.1.2 Factors Influencing Lending to SMEs

Most of the available literature has revealed several factors that influence lending both on the supply and demand side. Small and medium size enterprises lie on the demand side while commercial banks lie on the supply side. Bank size, credit risk, bank liquidity and the rate of interest have been found to affect the ability of commercial banks to advance funds to small and medium size enterprises.

Bank size is considered as one of the major determinant of bank lending decisions. Largercommercial banks provide a larger menu of financial services to customers as compared to smaller commercial banks. A study by Ladime et al (2013) established that capital structure and bank size have a statistically significant and positive relationship with bank lending behavior and that larger commercial banks tend to lend more to customers.

Additionally, lending to SMEs is considered riskier since they provide none or little collateral. According to the Financial Sector Deepening Kenya (2008), lack of cost-effective ways of credit risk quantification is one of the major reasons to why many lenders are reluctant to lend to SMEs. This is because, SMEs do not have standardized procedures and the available data may be of unreliable quality and accuracy. Studies point out that banks charge high interest premium for the borrowers who have higher credit default risk. Malede (2014) revealed that credit risk is a statistically significant determinant of commercial bank lending.

According to Laidroo (2012), liquidity and funding activity are highly correlated.Thus, commercial banks with a higherliquidity ratio are better protected as far as shocks to their deposit size (bank runs) is concern. This is an indication that they are able to lend more and are less vulnerable to economic shocks. Malede(2014)also established that liquidity ratio has a positive and statistically significant relation to commercial banks' lending decision, hence the higher the liquidity, the more commercial banks lends.

Interestrate also influences lending to all sectorsof economy. An increase in interest rate increases the cost of borrowing money.A study by Eze and Okoye (2014) established that, the lending rate is one of the major determinants of lending behavior by commercial banks in Nigeria of depositors' money and concluded that, it was necessary to reduce the lending rate ineconomyso as to encourage more borrowing for economic growth. However, Malede (2014) established that lending interest rate has a positiveand statistically insignificant relationship with commercial bank lending.

### 1.1.3 Commercial Banks in Kenya

Commercial banking sector is among the financial services sectors in Kenya that are expected to contribute greatly to the realization of country's Vision 2030. Commercial banks are very important source of funds for the operation and growth of businesses in Kenya.Lack of quality information about SMEs is one of the major SMEs-specific obstacles to SMEsfinancing. Family management hindrance is also another obstacle towards SMEs financing, since most SMEs in Kenya are family-owned. Further, the failure to standardize scoring models is another challenge, especially to those commercial banks, which have automated their SMEsfinancing systems (Calice et al. 2012).

Currently there are 43 registered commercial banks in Kenya, which are regulated by the Central Bank of Kenya. In the last decade, the bankingsector in Kenya has been characterized by increasing competition and innovation. This phenomenon has led to most commercial banks adopting innovative technology and creating more tailor made products in different sectors, especially the SME sector to improve the quality of their loan portfolio (Wambani, 2014).Most of commercial banks in Kenya have established separate units to specifically handle the needs of their SMEs customers, in acknowledgment of the inherent differences between SMEscustomers and corporate customers. Despite the risk involved in lending to SMEs, there is intense competition among Kenyan banks for SMEs customers. Some of the commercial banks in Kenya are providing trainings to their SME customersso as to improve their management and financial reporting skills (Calice et al., 2012).

**1.2 Research Problem**

The ability of SMEs to grow depends mostly on their ability to invest in restructuring, innovation and other factors that requirefunding. The access to fundingby SMEs is vitaltotheir growth and development. However, access to fundingremains one of the major challenges,especially to those SMEs in developing economies (Nkuah, 2013). To date, in most developing countries and Africa in particular, SMEs lack access to capital and money markets and still experience difficulties in obtaining capitaldespite efforts by some financial institutions and public sector bodies to open more avenues of funding (Kiama, 2012).

In addition, availability of external finance for SMEsis a topic of significant research interest to academic and an imperative issue to policy makers around the globe (Berger and Udell, 2005).The majority of the SMEsare still not considered credit worthy by commercial banks due to their inability to fulfill some conventional banking requirements (Alhassan and Sakara, 2014).As such, most SMEs in Kenyaare forced to consider other informal financing options, whose lending conditions are less stringent. The fundingobtainedfrom informal financing,is not enough to finance SMEs'expansion and growth. Therefore, it is important to identify the determinants of lending to SMEs by commercial banks in Kenya.

Several studies have been carried outbothinternationally and locally on the factors that influence lending to SMEs. For example, Haron et al (2013) examined the factors influencing SMEs in obtaining loans and established that collateral, good relationship with the lenders, and good financial records were some of the factors that influence lending to SMEs.Sun et al (2013) also established that SMEs financing confirm the severity of credit constraints to SMEs and that bank lending policy of using fixed assets as the security exacerbate the plight of SMEs funding. However, the studies did not examine the influence of credit risk, bank size, interest rate and liquidity of commercial banks on SMEs lending. In Kenya, a study by Langat (2013) examined the determinants of lending to farmers by commercial banks in Kenya and established that commercial banks give out loans to farmers that have reliable sources of income, but the study focused on the farming sector, hence its findings cannot be generalized to all SMEs.

Most studies on SMEs performance, growth and development acknowledge that lack of credit is the greatest constraint that SMEs face. Nonetheless, majority of the studies focus on the factors that influence the performance of SMEs and conclude that access to credit is utmost problem, which if solved can help mitigate the other factors.As such, most of these studies deviate from an in-depth analysis of the financial challenge facing SMEs. Instead, the studies give recommendations to SMEsand other stakeholdersonhow to mitigate or solvethefinancing problem without determining the factors that influence access to credit. Moreover, there is no comprehensive study in Kenyaon the determinants of lending to SMEs by commercial banks in Kenya. Thus, the aim of this study,whichintends to explore:What are the determinants of lending to SMEs by Commercial banks in Kenya?

**1.3 Research Objective**
Toestablishthe determinants of lending to SMEs by commercial banks in Kenya.

## II. Literature Review

Several authors have examined the various factors that influence lending to SMEs. For instance, Obamuyi (2007) examined the insight into the level of loan felony among the SMEs in Ondo State of Nigeria, and the lending practices of the country's banks towards SMEs. The study findings revealed that several factors were responsible for banks' altitude of not expanding loan portfolio, principal of which are poor credit worthiness, lack of collateral and other constraint imposed on banks' capital by regulations. Torre, Peria and Schmukler (2008)alsoassessedthe factors that commercial banks recognize as drivers and obstacles to SMEs financing. The study used a survey of commercial banks in Chile and Argentina. The study established that despite differences in the countries' environment, SMEs are strategic segment for most of the commercial banks in both Chile and Argentina.

In their study, Agyapong, Agyapong and Darfor (2011) analyzed the criteria for assessing SMEs' borrowers in Ghana. The study findings pointed out that when credit managers are deciding on whether to accept or reject a loan application from SMEs, the purpose of the loan, previous loans repayment history, repayment schedule, type of businessactivity involved, loan size relative to the business size and the provided collateral, ranked highest on their criteria list. The study revealed that lenders took meticulous interest in risk while dealing with SMEs. Hwarire (2012) also examinedcredit management and loan repayment of Small, micro and medium enterprises in a South African financial institution. The findings of the study revealed that 39 per cent of loan repayments by small, micro and medium enterprises were not made on time, while 28 per cent actually defaulted. In addition, the study established that gender, race and negative bank balance were found to be statistically significant in relation to default in credit management and loan repayment.

A study by Nguyen (2014) investigated the use of hard and soft information for commercial bank lending decisions to SMEs in Vietnam. The study findings revealed that although collateral based lending was the most used method and could substitute for other lending technologies, usually a combination of lending information types were used in their decision making process. Abdesamed and Wahab (2014) examined the financing of SMEs and the factors that influence SMEs to apply for a commercial bank loan. Using Logistic regression,the study found that business experience of business owners has no significant relation with the business's tendency to apply for a commercial bank loan. However, the study found that the educational background of the business owner,business size, collateral and loans with interest were negatively related to business tendency to apply for commercial bank loans.

In Kenya, Muchiti (2009) explored the institutional risk management strategies applied by commercial banks in Kenya while lending to SMEs and established that there were three distinctive institutional risk management strategies applied by commercial banks in Kenya when lending to SMEs borrowers. The strategies included risk-pooling strategies, the risk control strategies and risk avoidance strategies. Omboi and Wangai (2011) also analyzedthefactors that influencecredit demand among the small-scale entrepreneurs in Meru Central District, Kenya. The study established that the entrepreneur's household income, entrepreneur's level of education and number of entrepreneur's dependentsare significant factors that influence small-scale entrepreneurs to borrow from formal credit institutions.

In addition, Wachira (2011) investigated the factors that influence the use of microcredit amongst the SMEs based at Mutindwa market in Nairobi, Kenya and established that SMEs prefer Micro Finance Institutions loans because of their group-lending model, where security was by group guarantee demonstrating the fact that a majority of the loan consumers who are commonly women lack tangible collateral. Barasa(2013) also examined competition among lending financial institutions and easy accessibility to credit by SMEs in Nakuru, Kenya and found that Micro Finance Institutions and Savings and Credit Cooperative Societies were the most preferred sources of credit for SMEs. Finally, Kimutai and Jagongo (2013) explored the factors that influence credit rationing by commercial banks in Kenya. The study findings revealed that the major factors that influence credit rationing by commercial banks in Kenya are firm characteristics, loan characteristics and observable characteristics.

### III. Methodology

This study employed a descriptive research design. The study population consisted of the 43 commercial banks in Kenya. The study used secondary data from the annual published reports of commercial banks in Kenya,whichwasextractedfrom the statement of financial position and the statement of income as at 31$^{st}$December, 2014.The data covered a period of 5 years from 2010 - 2014. The data collected was analyzed through the multiple linear regression using the Statistical Package for Social Studies version 20. Regression analysis was used to analyze the relationship between Independent and Dependent variables. The regression equation took the following form

$$Y = \beta_0 + \beta_1 X_1 + \beta_2 X_2 + \beta_3 X_3 + \beta_4 X_4 + \varepsilon$$

Where;

$Y$ = Lending (measured using natural logarithm of net loans and advances to SMEs)
$X_1$ = Bank size (measured using natural logarithm of total assets)
$X_2$ = Credit risk (measured by non-performing loans/total assets)
$X_3$ = Liquidity ratio (measured by liquid assets/total deposits)
$X_4$ = Interest rate (measured by interest income/average loans)
$\beta_0$ = Constant
$\beta_1$ - $\beta_4$ = Regression coefficients
$\varepsilon$ = Error term

### IV. Data Analysis, Results and Discussion

#### 4.1 Descriptive Statistics

A census of the 43 commercial banks in Kenya was carried out but complete data was obtained from only 36 commercial banks hence a response rate of83.72%. The 83.72% response rate was considered adequate for the study.Table 1 shows the descriptive statistics from the research findings.

**Table 1:** Summary Statistics

|  | Minimum | Maximum | Mean | Std. Deviation |
|---|---|---|---|---|
| Lending to SMEs (log) | 5.80 | 8.45 | 7.1816 | .57450 |
| Bank Size (log) | 14.30 | 20.01 | 17.1903 | 1.30607 |
| Credit Risk (%) | 0.00 | 0.25 | .04087 | .040306 |
| Liquidity Ratio (%) | 21.00 | 112.80 | 41.6171 | 14.47975 |
| Interest Rate (%) | 13.00 | 24.00 | 17.3534 | 2.26210 |

**Source: Research Findings**

*Determinants of Lending to Small and Medium Enterprises by Commercial Banks in Kenya*The results in Table1, indicate that lending to SMEs had a mean of 7.18 and a standard deviation of 0.57 with minimum and maximum values of 5.80 and 8.45 respectively. Bank size had a mean of 17.19 and a standard deviation of 1.31 with minimum and maximum values of 14.30 and 20.01 respectively. Credit risk had a mean of 0.04 and standard deviation of 0.04 with minimum and maximum values of 0.00 and 0.25 respectively. Liquidity ratio had a mean of 41.62 and a standard deviation of 14.48 with minimum and maximum values of 21.00 and 112.80 respectively. Finally, interest rate had a mean of 17.35 and a standard deviation of 2.26 with minimum and maximum values of 13 and 24 respectively.

## 4.2 Correlation Analysis

**Table 2:** Correlation Matrix

|  | Lending to SMEs | Bank Size | Credit Risk | Liquidity Ratio | Interest Rate |
|---|---|---|---|---|---|
| Lending to SMEs | 1 |  |  |  |  |
| Bank Size | .972** | 1 |  |  |  |
| Credit Risk | -.360** | -.368** | 1 |  |  |
| Liquidity Ratio | -.303** | -.206** | -.076 | 1 |  |
| Interest Rate | .240** | .241** | -.137 | .068 | 1 |

*\*\*. Correlation is significant at the 0.01 level (2-tailed).*
**Source: Research Findings**

Table 2 shows that lending to SMEs has a strong and positive significant correlation with the size of the bank and a significant negative weak correlation with credit risk and liquidity. Lending to SMEs has a significant and weak positive correlation with interest rate.

## 4.3 Regression Analysis
### 4.3.1 Model Summary

**Table 3:** Model Summary

| Model | R | R Square | Adjusted R Square | Std. Error of the Estimate |
|---|---|---|---|---|
| 1 | .978a | .956 | .955 | .12172 |
| a. Predictors: (Constant), Interest Rate, Liquidity Ratio, Credit Risk, Bank Size ||||| 

**Source: Research Findings**

Table 3 shows that the coefficient of determination ($R^2$) = 0.956, which indicates that 95.6% of the variation in lending to SMEs is explained by the study variables. The 4.4% of the variation is explained by other factors outside the model and the error term. The correlation coefficient of 0.978 indicates that there is a strong positive relationship between lending to SMEs and the determinants of lending to SMEs by commercial banks in Kenya.

### 4.3.2 Anova

**Table 4:** Anova

| Model |  | Sum of Squares | df | Mean Square | F | Sig. |
|---|---|---|---|---|---|---|
| 1 | Regression | 56.487 | 4 | 14.122 | 953.218 | .000b |
|  | Residual | 2.593 | 175 | .015 |  |  |
|  | Total | 59.079 | 179 |  |  |  |
| a. Dependent Variable: Lending to SMEs |||||||
| b. Predictors: (Constant), Interest Rate, Liquidity Ratio, Credit Risk, Bank Size |||||||

**Source: Research Findings**

Table 4 indicates that the f value of 953.218 is significant at 95% confidence level since the p-value (0.000<0.05), an indicator that the model was a good fit. Additionally, the regression sum of squares (56.487) is more than residual, indicating that the study variables explain a greater proportion of the variation in the model.

### 4.3.3 Regression Coefficients

**Table 5:** Regression Coefficients

| Model |  | Unstandardized Coefficients |  | Standardized Coefficients | t | Sig. |
|---|---|---|---|---|---|---|
|  |  | B | Std. Error | Beta |  |  |
| 1 | (Constant) | .222 | .152 |  | 1.465 | .145 |
|  | Bank Size | .412 | .008 | .936 | 51.733 | .000 |
|  | Credit Risk | -.312 | .246 | -.022 | -1.265 | .208 |
|  | Liquidity Ratio | -.005 | .001 | -.114 | -6.882 | .000 |
|  | Interest Rate | .005 | .004 | .019 | 1.177 | .241 |
| a. Dependent Variable: Lending to SMEs |||||||

**Source: Research Findings**

As per the study results, the following regression equation can be specified:
$$Y = 0.222 + 0.412X_1 - 0.312X_2 - 0.005X_3 + 0.005X_4 + \varepsilon$$

Table 5 shows that banksizehas a significant positive relationship with lending to SMEs by commercial banks in Kenya while credit risk has a negativeinsignificant relationship. In addition, the findings show that liquidity has a significant negative relationship with lending to SMEs while interest rate has a positive insignificant relationship with lending. These findings indicate that bank size and liquidity significantly influences lending to SMEs while credit risk and interest rate do not have significant influence on lending to SMEs by Commercial banks in Kenya.

**4.4 Interpretation of the Findings**
The study established that bank size and liquidity significantly influences lending to SMEs by commercial banks in Kenya. In addition, the study established that credit risk and interest rate have no significant influence on lending to SMEs by commercial banks in Kenya. These findings are similar to those of Ladime et al (2013) who establishedthatbank size and capital structure have a statistically significant and positive relationship with bank lending behavior and that, bigger banks seem to be in a better position to lend more.Malede (2014) also established that bank size has a positive and statistically significant influence on commercial bank lending. On interest rate, the findings conform to that of Malede (2014) who found that the lending interest rate has a positivebutstatistically insignificant relationship with commercial banks' lending.

## V. Conclusion

The aim of this study was to explorethe determinants of lending to SMEs by commercial banks in Kenya. The study concludes that bank size has a significant positive relationship with commercial bank lending to SMEs by commercial banks in Kenya. However, liquidity is negatively, significantly related to lending. The study also concludes credit risk and interest rate do not have significant influence on lending to SMEs by commercial banks in Kenya. Based on the study findings, the study recommends that effective policies should be developed to ensure commercial banks grow and therefore advance more credit to SMEs.